# Metrics of Risk Associated with Defects Rediscovery

A.V. Miranskyy[a], M. Davison[b], M. Reesor[c]
[a]IBM Canada Ltd., Toronto, ON, Canada
Department of { [a,b,c]Applied Mathematics | [b]Statistical & Actuarial Sciences | [b]Richard Ivey School of Business}, University of Western Ontario, Canada
[a]andriy@ca.ibm.com, { [b]mdavison | [c]mreesor}@uwo.ca

## Abstract

Software defects rediscovered by a large number of customers affect various stakeholders and may: 1) hint at gaps in a software manufacturer's Quality Assurance (QA) processes, 2) lead to an overload of a software manufacturer's support and maintenance teams, and 3) consume customers' resources, leading to a loss of reputation and a decrease in sales.

Quantifying risk associated with the rediscovery of defects can help all of these stakeholders. In this chapter we present a set of metrics needed to quantify the risks. The metrics are designed to help: 1) the QA team to assess their processes; 2) the support and maintenance teams to allocate their resources; and 3) the customers to assess the risk associated with using the software product. The paper includes a validation case study which applies the risk metrics to industrial data. To calculate the metrics we use mathematical instruments like the heavy-tailed Kappa distribution and the G/M/k queuing model.

## 1 Introduction

During in-house testing of a software product, the Quality Assurance (QA) team attempts to remove defects injected during software development. It is impossible to remove all defects before shipping the software product. As customers use the product, they *discover* defects that "escaped" the QA team. After receiving a defect report the software provider's maintenance team prepares and makes available a fix. A discovered defect is sometimes *rediscovered* by another customer. This rediscovery could occur because another customer finds the defect before the fix is available or has not been installed. A defects relating to rarely used software features will be rediscovered infrequently. However, a defect relating to popular and extensively used features may affect a significant fraction of customers.

Frequently rediscovered defects (affecting many customers) can cause an *avalanche* of requests, defined as a large number of requests for fixes from multiple customers within a short timeframe. Because different software versions are run on different software and hardware platforms, each customer may require a different fix.

An avalanche has significant consequences: support personnel will experience a heavy volume of support requests and the maintenance team will need to prepare a large number of special builds while customers await the official fix. On the other side, the customers' system administrators will need to spend time assessing the fix's risk and distributing it to their systems. An inordinate number of defects may diminish the provider's reputation and result in decreased software sales.

Frequent rediscovery of a defect suggests that one or more common functionalities were not properly tested. Analysis of such defects is important to identify gaps in QA processes to prevent the future escape of similar defects.

Defect risk analysis is therefore important for software manufacturers and customers alike. We propose a set of quantitative risk metrics which can be used to assist:

- The support team's assessment of the potential number of repeated calls on the same subject, helping in personnel allocation;
- The maintenance team's estimation of the potential number of repeated special builds,

assisting in resource allocation of team members;
- The QA team's assessment of trends in frequently rediscovered defects on release-to-release basis. If the trend shows increased defect rediscovery, QA processes must be improved. The resulting strategy to close testing process gaps can be derived by root cause analysis of frequently rediscovered defects;
- The customer's assessment of the risks associated with software product use.

We present a validation case study showing applicability of these new metrics to an industrial dataset of defects rediscovery. In order to model the data we derive a compound Kappa distribution and use the G/M/k queueing model.

Section 2 of this paper reviews relevant work. Motivation for the metrics applications is described in Section 3 with their formal definitions deferred until Section 4. Section 5 provides a validation case study, showing application of the metrics to the industrial data. Finally, Section 6 concludes the paper.

## 2  Related Research

The paper's main contribution is a set of metrics for assessing defect rediscovery risks. The following metrics have been formulated by other authors: the number of rediscoveries per defect [1], the time interval between first and last rediscovery of a given defect[1] and the probability that a customer will observe failure in a given timeframe [2]. Our metrics are complementary to these three. Our metrics can help in resource allocation of service and maintenance teams; these metrics rely on information about arrival of defect rediscoveries. Other authors have used counting processes [3] and regression models to help estimate staffing needs. However, the authors do not assess risks associated with under-staffing; hence our work complements theirs.

We use a G/M/k queue analysis to estimate staffing needs for delivery of special builds fixing rediscoveries for customers. Queuing theory tools have not yet been applied to this problem, although the load on a k-member service team delivering fixes for initial rediscovery of defects was modeled in [4] using $k$ parallel M/M/1 processes. Work has also been done on modeling the initial discovery repair time distribution [5] and predicting defect repair time based on attributes of past defect reports [6].

The second contribution of this paper is the introduction of a compound Kappa distribution, related to the family of heavy-tailed distributions, to model the data. While previous work has observed that, depending on the dataset, distribution of defect rediscoveries have found either to be thin-tailed (exponentially bounded) [7] or, in other cases, heavy-tailed [8], [9], many processes in software engineering are governed by heavy-tailed distributions [10]. Based on these observations, modeling the rediscovery distribution was performed using the empirical [8], geometric [7], lognormal [9], and Pareto [9] distributions. We found that none of these parametric models provided an adequate fit to our data. Therefore, we introduced the more flexible compound Kappa distribution for the number of rediscoveries providing reasonable fit.

## 3  What Do We Want in a Risk Metric?

Metrics used by Support and Maintenance Teams, Quality Assurance Team and customers are given in Sections 3.1, 3.2 and 3.3, respectively.

### 3.1  Support and Maintenance Teams

Defect discovery related to common and frequently executed functionalities triggers a large number of support requests shortly after its initial discovery. This can be explained as follows:

*Proactive requests for software fix*: The software manufacturer publishes information about newly discovered defects on a regular basis. In turn, a customer's software administrators analyze newly published defects shortly after publication and use their expertise to assess the defect rediscovery probability and the severity of implications associated with its rediscovery. If the administrators decide the risks warrant it, they will contact the manufacturer's support desk requesting a special software build[1] incorporating a defect fix. This is a preventative measure against encountering this problem in the future.

---
[1] We assume that the standard vehicle for delivery of fixes is through cumulative fix packs.

*Reactive requests for software fix*: A customer could encounter a defect recently exposed by another customer. This is common for regression defects which break existing functionality, thus, requesting a special build from the support desk to prevent defect reencounter.

In both cases, the support desk will, after an initial assessment, relay this special build compiling and testing request to the manufacturer's maintenance team. Large numbers of customers classifying a defect as "potentially discoverable" may trigger an avalanche of special build requests. These requests can overload the maintenance and support personnel. We now analyze the cause of the overload and the actions needed to prevent it.

**Maintenance Team:** Customers may use different versions of the product on multiple platforms. Even though the source code repairing a given defect is the same, special builds will have to be tailored individually for each customer. Building and testing a special build of a large software product can take several days, consuming human and hardware resources. Therefore, the maintenance team is interested in knowing the probability of the increase in the number of requests for special builds above a set threshold[2] over a particular time interval as well as the total number of requests above the threshold. When the number of requests for special builds exceeds a certain threshold we say a "*spike*" results. In addition to the probability of a spike, the maintenance team is also interested in the conditional expectation of the spike's size given its occurrence. Also of interest is the probability that the number of requests for special builds in a given timeframe will not exceed a predetermined threshold. By leveraging this data, the management of the maintenance team can allocate personnel (based on the expected number of special builds and the average waiting time to deliver the builds) so that they can be transferred to the "special-build team" on an as-needed basis, decreasing delivery time to customer.

**Support Team**: Once contacted by a customer with a proactive request for a special build, a support analyst must verify if the defect of interest can be rediscovered by this customer[3]. If the request is reactive, then the analyst has to verify that the problem is caused by this particular defect and not another one with similar symptoms. Knowing the probability and potential size of spikes in the number of requests (as well as the probability of not exceeding a certain number of calls in a given timeframe) can support management's personnel allocation, speeding diagnostics thus leading to faster transfers to the maintenance team and a decrease in the overall turnaround time. The end result is cost savings and higher customer satisfaction.

## 3.2 Quality Assurance Team

Maintenance and Support teams can use information about frequently rediscovered defects for tactical planning. The QA team can use this data for strategic planning to identify trends in software quality on a release-to-release basis. Frequently rediscovered defects affecting a significant fraction of the customer base relates to frequently executed common functionality. The presence of such defects suggests the QA team's inability to reproduce customer workloads in-house or its failure to execute existing test-cases covering this functionality [11]. In order to compare releases of the product, an analyst needs to find out how many defects were rediscovered at least $x$ times for a given release[4]. The numbers of defects with high number of rediscoveries should decrease from release to release. An increasing number of defects may imply a deterioration of QA processes. The QA team should analyze root causes of defects to find the actions needed to close these gaps.

## 3.3 Customers

Information about defect rediscovery interests customers, especially for mission-critical applications. It is known that customers perceive quality [11-13] to be correlated with the quantity and severity of failures they encounter. Therefore, comparison of the number of defects affecting a significant percentage of the customer base for various products can be used as one of the measures needed to select the "safest" product. In the next section we discuss some techniques required to answer these questions.

---

[2] In addition to routine requests for special builds for defects with small numbers of rediscoveries.

[3] For example, even though the customer is using functionality affected by a given defect, the problem could be specific to a hardware platform not used by this customer.

[4] If the customer base of a software product does not change significantly from release to release, then the number of defects with a high number of rediscoveries can be directly compared. If this assumption fails then it may be beneficial to normalize the number of defects by the size of customer base and/or product usage.

# 4 Definition of New Risk Metrics

Based on the discussion in the previous section, stakeholders are interested in the following data:
1. The number of defects rediscovered more than certain number of times in a given timeframe;
2. The number of defects affecting a certain percentage of the customer base in a given timeframe;
3. The total number of rediscoveries for defects rediscovered more than certain number of times in a given timeframe;
4. The probability of spikes in the number of requests in a given timeframe;
5. The probability that the number of requests for special builds in a given timeframe will not exceed a certain threshold;
6. The worst-case scenario for the total number of rediscoveries;
7. The expected waiting time of customers.

To calculate these variables we build a formal statistical model of defect rediscoveries.

Suppose that $N$ field defects are discovered independently[5] up to time $t$ with the $i$-th defect rediscovered $D_i \equiv D_i(s,t)$ times in the interval $[s, t)$, $s<t$. For the sake of brevity we will use $D_i$ and $D_i(s,t)$ interchangeably. The total number of rediscoveries $U(s,t)$ between times $s$ and $t$ is given by

$$U(s,t) \equiv \sum_{i=1}^{N(t)} D_i(s,t). \quad (1)$$

A spike (of strength $r$) is defined to occur when the total number of rediscoveries in a given timeframe $[s, t]$ is greater than $r$:

$$U(s,t) > r. \quad (2)$$

The probability that the $i$-th defect will be rediscovered exactly $d$ times in the interval $[s,t)$ is given by $p_i(d) \equiv P(D_i = d)$. We assume that the probability distribution of the number of rediscoveries is the same for all defects (i.e., that the $D_i$ are identically distributed random variables). The same assumption is implicitly used in [7-9]. This assumption is not necessary to compute the metrics; however, it simplifies calibration of our model[6].

Assuming that the number of rediscoveries lies in the range $[0,\infty)$, the probability that the number of rediscoveries of the $i$-th defect will be less than or equal to $d$ is given by cumulative distribution function (cdf)

$$F_i(d) = E\left[I_{D_i \leq d}\right] = P(D_i \leq d) = \sum_{j=0}^{d} p_i(j), \quad (3)$$

where $I_A$ is an indicator variable such that

$$I_A = \begin{cases} 1, & \text{if } A \text{ holds} \\ 0, & \text{otherwise} \end{cases}; \quad (4)$$

from which it follows that the expected value of $I_A$ is equal to probability of $A$:

$$E[I_A] = P(A). \quad (5)$$

The probability that the number of rediscoveries of the $i$-th defect will be greater than $d$ is given by the decumulative distribution function:

$$\tilde{F}_i(d) = E\left[I_{D_i > d}\right] = P(D_i > d)$$
$$= \sum_{j=d+1}^{\infty} p_i(j) = 1 - F_i(d). \quad (6)$$

The quantile function, (inverse of the cdf) $F_i^{-1}(\alpha)$ is used to determine the $\alpha$ quantile of a given distribution.

The expected total number of rediscoveries for the $i$-th defect with rediscoveries ranging between $l$ and $u$ is given by

$$R_i(l,u) = E\left[D_i I_{l \leq D_i \leq u}\right] = \sum_{j=l}^{u} j p_i(j). \quad (7)$$

Note that $R_i(1,\infty)$ calculates expected number of rediscoveries of the $i$-th defect. Armed with these instruments, we can estimate the metrics needed to address the questions listed above.

---

[5] Initial exposure of a defect depends only on user's workload and configuration. Discovery of one defect should not, in the majority of cases, change probability of discovery of another defect.

[6] If, in a particular case, defects are not identically distributed, an analyst can try to partition the dataset into subsets of identically distributed defects based on a classification criterion.

**M$_1$: Expected number of defects rediscovered more than certain number of times**

The expected number of defects rediscovered more than $d$ times is given by

$$M_1(d) = E\left[\sum_{i=1}^{N} I_{D_i > d}\right]$$
$$= \sum_{i=1}^{N} E\left[I_{D_i > d}\right] = \sum_{i=1}^{N} \tilde{F}_i(d). \quad (8)$$

If all $D_i$ are identically distributed, then (8) simplifies to

$$M_1(d) = \sum_{i=1}^{N} \tilde{F}_i(d) \stackrel{i.d.}{=} N\tilde{F}_1(d) = N\tilde{F}(d). \quad (9)$$

Note that we suppress indices to ease notation.

**M$_2$: Expected number of defects affecting certain percentage of the customer base**

This metric is similar to M$_1$. If we denote the total number of customers by $C$ and assume that every customer rediscovers a given defect only once, then the relation between the percentage of the customer base $x$ and number of rediscoveries $d$ is given by

$$\tilde{d} \approx \lfloor xC/100 \rfloor, \quad (10)$$

where $\lfloor \cdot \rfloor$ is the floor function mapping to the next smallest integer. M$_2$ is calculated as

$$M_2(\tilde{d}) = \sum_{i=1}^{N} F_i(\tilde{d}) \stackrel{i.d.}{=} NF_1(\tilde{d}) = NF(\tilde{d}), \quad (11)$$

**M$_3$: Expected total number of rediscoveries for defects with number of rediscoveries above certain threshold in a given timeframe**

The expected total number of rediscoveries for a given spike is calculated as

$$M_3(d) = E\left[\sum_{i=1}^{N} D_i I_{d \leq D_i < \infty}\right]$$
$$= \sum_{i=1}^{N} E\left[D_i I_{d \leq D_i < \infty}\right] = \sum_{i=1}^{N} R_i(d, \infty) \quad (12)$$
$$\stackrel{i.d.}{=} NR_1(d, \infty) = NR(d, \infty),$$

where $d$ is the smallest number of rediscoveries of a particular defect.

**M$_4$: Probability of spikes in the number of requests in a given timeframe**

This can be rephrased as probability that the total number of rediscoveries will exceed a certain threshold $L$. The calculation of this value involves two steps:

Find $d$ to satisfy the equation:

$$L = E\left[\sum_{i=1}^{N} D_i I_{1 \leq D_i \leq d}\right]$$
$$= \sum_{i=1}^{N} E\left[D_i I_{1 \leq D_i \leq d}\right] = \sum_{i=1}^{N} R_i(1, d) \quad (13)$$
$$\stackrel{i.d.}{=} NR_1(1, d) = NR(1, d),$$

Since $d$ is discrete, we will not always be able to find an integer value of $d$ to satisfy this equality, so we look for the smallest integer $d$ which satisfies:

$$L \leq \sum_{i=1}^{N} R_i(1, d) \stackrel{i.d.}{=} NR_1(1, d) = NR(1, d), \quad (14)$$

After identifying $d$, the probability that the total number of rediscoveries will exceed $L$ is given by

$$M_4(d) = \tilde{F}(d) = 1 - F(d). \quad (15)$$

**M$_5$: Probability that the total number of rediscoveries will not exceed certain threshold**

This metric is complementary to M$_4$ and is calculated in a similar manner. Given the number of rediscoveries $d$ from (14) we calculate M$_5$ as

$$M_5(d) = 1 - M_4(d) = F(d). \quad (16)$$

**M$_6$: Estimate of the worst case scenario for the total number of rediscoveries**

This metric provides a threshold which the total number of rediscoveries will not exceed for a given probability level. The metric provides the worst case scenario of the total number of rediscoveries. For example, if the value of M$_6$(0.99) is equal to $y$, then it will tell us that in 99 cases out of 100 the total number of rediscoveries will not exceed $y$[7].

In order to obtain this value we need to identify number of rediscoveries for a given probability level $\alpha$ using $\lfloor F_i^{-1}(\alpha) \rfloor$. The threshold value of rediscoveries is then calculated using

---

[7] This is similar to "Value At Risk" measure used in finance.

$$M_6(\alpha) = E\left[\sum_{i=1}^{N} D_i I_{1 \le D_i \le \lfloor F_i^{-1}(\alpha) \rfloor}\right]$$

$$= \sum_{i=1}^{N} E\left[D_i I_{1 \le D_i \le \lfloor F_i^{-1}(\alpha) \rfloor}\right]$$

$$= \sum_{i=1}^{N} R_i\left[1, \lfloor F_i^{-1}(\alpha) \rfloor\right] \quad (17)$$

$$\stackrel{i.d.}{=} N R_1\left[1, \lfloor F_1^{-1}(\alpha) \rfloor\right]$$

$$= N R\left[1, \lfloor F^{-1}(\alpha) \rfloor\right].$$

**$M_7$: Expected waiting time of customers being serviced**

This metric is calculated using queuing tools [14]. $M_7$ depends on the distributions governing service time, requests' inter-arrival time and number of personnel allocated to handle these requests. The metric's formula will depend on the form of distributions governing the queue.

## 4.1 Application of the Risk Metrics

Metrics $M_1$ and $M_2$ can be used by QA and customers to calculate the number of defects injected in common functionality ($M_1$) and identify defects affecting a certain fraction of the customer base ($M_2$) as discussed in Sections 3.2 and 3.3.

Metric $M_3$ helps to estimate the total number of rediscoveries for frequently discovered defects and the potential contribution of the frequently rediscovered defects to the overall load of support and maintenance teams.

Metrics $M_{4-7}$ can be used to address issues described in Section 3.1 and help in resource allocation of the service and maintenance teams.

Metrics $M_{4-6}$ may also be used for resource allocation as follows: a manager responsible for resource allocation knows the number of available personnel, denoted by $A$, and, based on historical data, the average quantity of service (special build) requests that support (maintenance) person can process per unit time, denoted by $\mu$.

A simple estimate[8] of the overall amount of service (special build) requests, denoted by $Q$, that can be processed by personnel in a given period $T$ is given by

$$Q = A\mu T. \quad (18)$$

The manager can then use $M_4(Q)$ or $M_5(Q)$ to get an estimate of the probability that the support of maintenance team will be able to handle the volume of requests $Q$.

The manager can examine the resource allocation task from the opposite perspective: instead of calculating the probability of handling requests by employees, she can calculate the number of service or special build requests that will not exceed $M_6(\alpha)$ at confidence level $\alpha$. We can obtain the amount of personnel $A$ needed to handle this workload by inverting (18):

$$A = M_6(\alpha) / (\mu T). \quad (19)$$

Note that stationary processes should be used if metrics $M_{4-7}$ are used for forecast-related management decisions. In order to calculate the metrics $M_{1-6}$ we also need an estimate of the total number of defects. There exists a variety of methods that can be used to estimate this value as reviewed in [15]. Detailed discussion of these techniques is beyond the scope of this paper.

The metrics can be used in-process by selecting defects within a timeframe of interest[9]. For example, one can compare expected number of defects, $M_1$, for the latest and previous releases based on the information available six months after the shipping of each release. Another example is estimation of the expected waiting time of customers, $M_7$, in the next three months.

## 5 Case Study

In this case study we use defect discovery data for a set of components of four consecutive releases of a large scale enterprise software. To preserve data confidentiality, the dataset is scaled and rounded. For the same reason, we assume that the customer base size remains constant across all four releases[10].

Figure 1 depicts $N(t)$, the cumulative number of defects encountered up to $t$ years after the product has shipped. The total number of rediscoveries from time 0 (general availability (GA) date of the product to be shipped to the field) to time $t$, $R(0,t)$, is shown in Figure 2. The age of the releases in

---

[8] This estimate does not account for request inter-arrival times. A better estimate can be obtained using metric $M_7$.

[9] Some historical data is required for calibration.

[10] This will affect estimation of the $M_2$. However, the size of the customer base acts as normalization factor and should not affect comprehension of the metric application.

the field varies from 5 years for v.1 to 2 years for v.4 because v.4 was released about three years after v.1

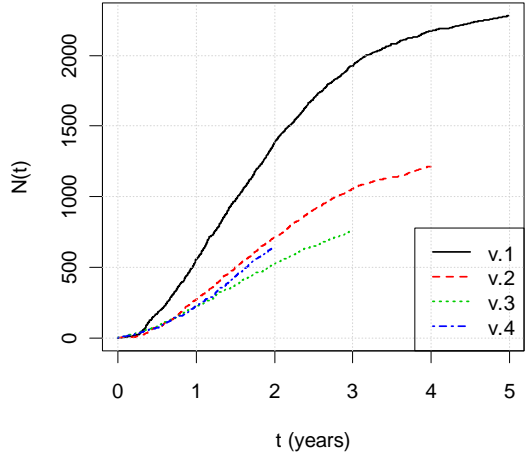

Figure 1. $N(t)$: total number of defects discovered up to time $t$.

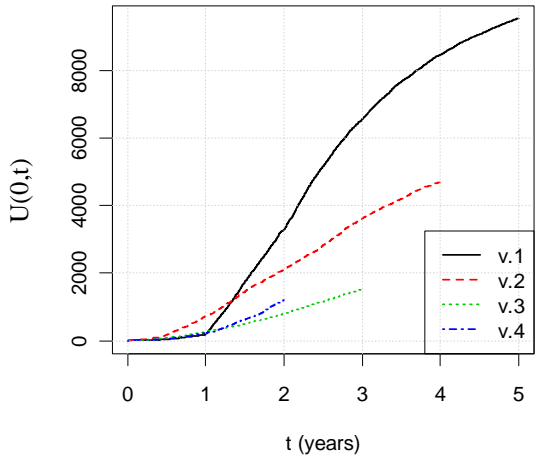

Figure 2. $U(0,t)$: total number of rediscoveries up to time $t$.

Metrics $M_{1-6}$ rely on the distribution of the number of rediscoveries per defect $D_i$ (Section 4). In the same section, to simplify formulas for $M_{1-6}$, we assumed that $D_i$ are identically distributed so must specify the distribution of $D_i$. Without loss of generality, we split the $D_i$ data for every release into yearly time intervals $D_i(t, t+1)$, where $t = 0 \ldots 4$ (if the data is present for a given release).

This split is reasonable in practice. Resource planning (metrics $M_{3-6}$) is performed for a fairly short future time interval; one year or less being common planning horizons. Metrics $M_{1-2}$ focus on measuring general quality of the product and would benefit from information about rediscoveries over the complete lifecycle of a product in the field. However, it is also critical to identify issues with QA processes early, so that actions can be taken to improve QA processes of releases under development. Since the lifespan of an enterprise software product can often reach a decade or more, it would not be practical to wait such a long time to obtain information.

## 5.1 Finding a Suitable Distribution

In order to find an adequate distribution for a given dataset an analyst should evaluate the goodness of fit for each potential distribution using such techniques as a QQ-plot and an L-moments ratio diagram [16]. Once a subset of candidates is identified, an analyst can use Akaike's information criterion (AIC) [17] to select the best distribution from the subset (by balancing distribution complexity and goodness of fit). The details of the distributional analysis for our dataset are given in Appendix A. According to the AIC, the compound Kappa distribution provides the best fit for three out of four yearly datasets.

## 5.2 Application of the Metrics

The application section is divided into two parts. Section 4.2.1 focuses on software quality metrics, while Section 4.2.2 concentrates on resource allocation-related metrics.

### 5.2.1 Analysis of Software Quality

As discussed in Sections 3. and 4, metrics $M_1$, defined by equation (9), and $M_2$, equation (11), can be used to identify potential issues with QA processes and to help customers find the "safest" product. Figure 3 plots $M_1$ against the number of rediscoveries $d$. The plot shows that from v.1 to v.3 the number of defects rediscovered more than d times decreased for all values of d. However, the value of the metric went up for v.4:

$$M_1^{v.1} < M_1^{v.2} < M_1^{v.3} < M_1^{v.4}.$$

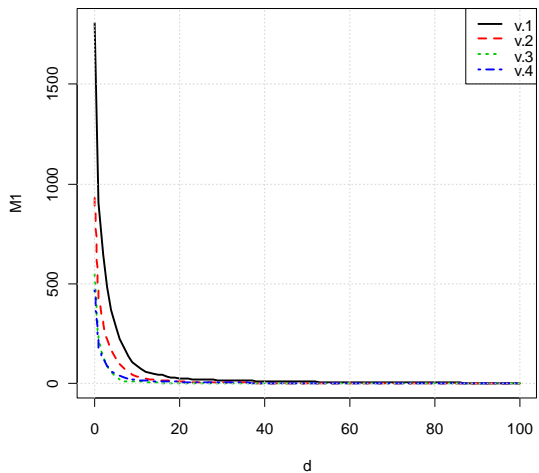

Figure 3. $M_1$: expected number of defects rediscovered more than *d* times during the 2$^{nd}$ year after GA date.

This ordering becomes especially obvious if we look at values of $M_1$ for a specific number of rediscoveries (for example, *d*=10)[11] for all releases (shown in Table 1).

Table 1. Values of variables

| Variable | v.1 | v.2 | v.3 | v.4 |
|---|---|---|---|---|
| $M_1(10)$ | 87.08 | 35.26 | 7.88 | 19.00 |
| R(1,2)/N(2) | 2.24 | 1.76 | 1.02 | 1.53 |

This information suggests that the quality $\mathbb{Q}$ of QA processes went down in the last release:
$$\mathbb{Q}(v.1) < \mathbb{Q}(v.2) < \mathbb{Q}(v.4) < \mathbb{Q}(v.3). \qquad (20)$$
Before making this conclusion we should look at other quality attributes of the software. The number of rediscoveries per defect: $U(1,2) / N(2)$, given in Table 1; the ordering of the total number of defects (Figure 1), and their rediscoveries (Figure 2) in the second year concur our hypothesis (20).

Based on this conclusion, an analyst needs to identify gaps in QA processes by analyzing reasons for the defects' injection and subsequent escape to the field. Upon identifying the gaps, actions should be derived and taken to prevent injection of defects in future software releases. Additional data can be extracted by focusing the analysis on subsets of data grouped by testing team, functionality, etc.

Since we assume that the number of customers remains constant for all four releases Eq. (10) implies metric $M_2$ is a scaled version of $M_1$. Therefore, the number of defects affecting a certain fraction of the customer base is larger for v.4 than for v.3. At this stage a customer should perform a risk-benefit analysis: would the value of v.4's new features outweigh the increased risk of encountering defects[12]. The customer can perform additional analysis by looking at $M_2$ for a specific subset of defects that may critically affect operations, e.g., defects in critical functionality leading to a software crash, while omitting defects that are related to functionality not used by this particular customer.

### 5.2.2   Resource Allocation

The application of metrics $M_{3-7}$ for resource allocation was discussed in Sections 3.1 and 4.

A few examples of the value of these metrics are as follows. Suppose that the maintenance team manager needs to analyze recourse allocation for building rediscovery-related special builds for v.4 during the third year of service. Currently, the manager has *k*=8 people allocated to this task. Given an available fix, the manager knows that the average time for a team member to create, test, and ship a special build is two days: a person can handle on average $\mu = 250 / 2 = 125$ requests per year[13]. Based on historical data, we know that the process governing the arrival of rediscoveries during the third year is the same as during the second year. Therefore, we can use the data from the second year to get resource allocation estimates for the third year.

---

[11] We pick this number arbitrarily; an analyst can pick this threshold value based on their expertise on problematic levels of rediscoveries in their organization.

[12] The complete analysis should include additional factors, such as software cost and support lifespan
[13] Assuming 250 working days per year.

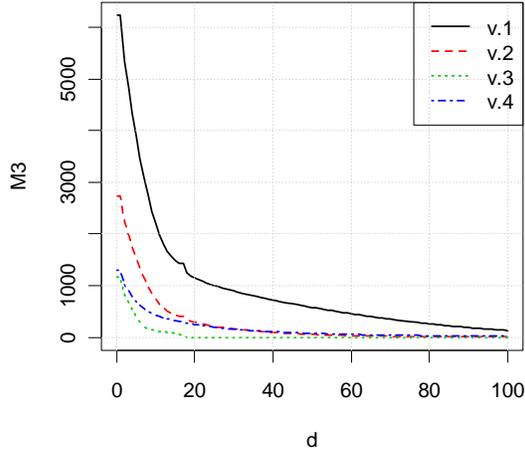

Figure 4. $M_3$: expected total number of rediscoveries for defects with number of rediscoveries above $d$ during the 2$^{nd}$ year after GA date.

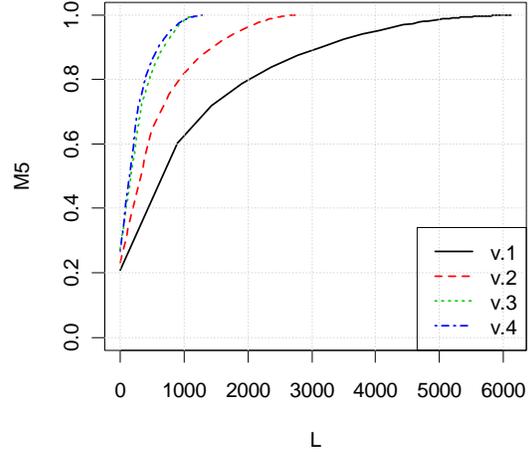

Figure 5. $M_5$: probability that the total number of rediscoveries will not exceed $L$ during the 2$^{nd}$ year after GA date.

To put the importance of this team into perspective, the manager needs to know the fraction of rediscovery-related special builds compared to the total number of requests for special builds. The total expected number of rediscoveries is given by $M_3(1)$ (Eq. (12)) and shown in Figure 4. For simplicity, we use the number of defects discovered during the second year as an estimate of the number of discoveries during the third year. In this case (based on Figure 1) the expected number of discovered defects during the third year is approximately equal to $N(2) - N(1) = 417$. $M_3(1)$ for v.4 is equal to 1299. The fraction of the total number of request for special builds related to rediscovered defects is $1299 / (1299 + 417) \approx 0.76$. This team will handle a significant portion of the overall number of requests and, therefore, allocation of the personnel for this team can be critical.

Equation (18) can be used to get the average number of requests that the team can handle per year:

$$Q = k\mu T = 8 \times 125 \times 1 = 1000. \qquad (21)$$

Metric $M_5(Q)$ (Eq. (16)) is the probability that the number of requests will not exceed $Q$. Based on Figure 5, $M_5(Q) = M_5(1000) \approx 0.984$. This value can be interpreted as follows: in the hypothetical case of the software being in service for 1000 years (and the arrival of rediscoveries being stationary) in 984 years out of 1000 a team of 8 people would be able to handle the requests, in 16 years out of 1000 the number of requests would be larger than this team can handle.

What if the manager would like to know how many people are needed to handle requests in 999 years out of 1000? By using Equation (19) and Figure 6, the number of people needed to handle these requests is equal to:

$$k = \frac{M_6(\alpha)}{\mu T} = \frac{M_6(0.999)}{125 \times 1} \approx \frac{1245}{125} \approx 10. \qquad (22)$$

This suggests that the manager should allocate two additional team members to handle this extreme case.

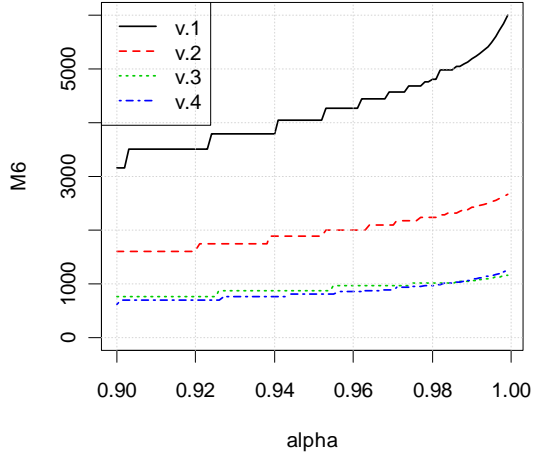

Figure 6. Estimate that the total number of rediscoveries will not exceed $M_6$ with confidence level $\alpha$.

So far we have not considered the amount of time customers must wait to get their special build. If, at a certain time, the maintenance team receives an avalanche of requests, the customers will have to wait for a long time to obtain their special builds. In order to obtain the expected waiting time, $W$, we need to model this queue [15] to consider 1) the distribution of request inter-arrival times. [14]; and 2) the time to complete service. We assume that the process is stationary, the queue is "First in, First out", and the service times are exponentially distributed with mean service time equal to $1/\mu = 1/125 = 0.008$ years. The empirical average number of requests for special builds of v.4 during the second year is $\lambda = 982$ requests per year. The distribution of inter-arrival times for v.4 (second year[14]) is given in Figure 7. We could not find an analytic distribution providing a good fit to the data. Due to this fact, we pick a queuing model denoted, using Kendall's notation [14], as G/M/k:

- G: **g**eneral distribution of inter-arrival requests. In our case we will use the empirical distribution in Figure 7,
- M: exponential distribution of service times,
- k: number of team members handling the requests.

Details of the model are given in [14].

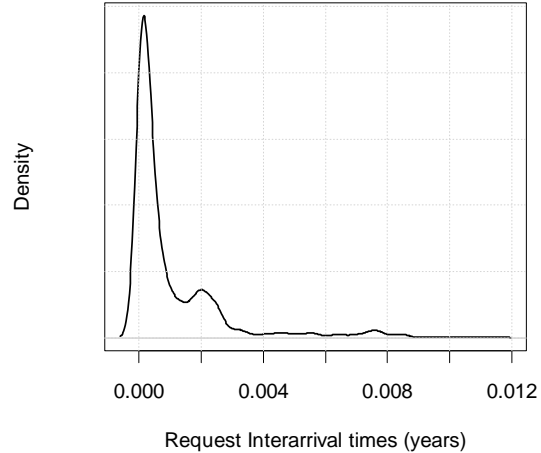

Figure 7. Density of requests inter-arrival times for v.4, second year.

Complementary to $W$, we calculate the percentage of the overall working time the team members spend generating special builds:

$$b = \lambda / (k\mu) \times 100. \qquad (23)$$

For example, average busy time for 8 team members is

$$b = 982 / (8 \times 125) \times 100 \approx 98.2\%.$$

Model results are given in Table 2. Note that the table shows waiting time in the queue $W_q$; in order to obtain the average waiting time $W$ we need to add average waiting time $\mu$ (2 days) to $W_q$. The table shows that 8 team members can handle service requests. However, average waiting time will be 26.3 working days, which may be unacceptably long. Increasing the team to 10 decreases $W$ to 2.9 days and a 12 member team further reduces $W$ to 2.2 days. However the associated busy time of team members drops from 98.2% for 8 team members to 65.5% for 12 team members.

With this information, the manager can now select the optimal team size and plan additional tasks for the team members to fill their free time. The analysis of support personnel allocation is performed in a similar manner.

---

[14] As before, we assume that that the process governing the arrival of rediscoveries during the third year is the same as during the second year.

Table 2. Results of the G/M/k model for v.4, second year.

| Number of team members k | Expected waiting time in the queue $W_q$ in working days | | Percent of the time the team members are busy ($b$) |
|---|---|---|---|
| | G/M/k | M/M/k | |
| 8 | 24.31 | 13.09 | 98.2% |
| 9 | 2.26 | 1.07 | 87.3% |
| 10 | 0.88 | 0.35 | 78.6% |
| 11 | 0.43 | 0.14 | 71.4% |
| 12 | 0.23 | 0.06 | 65.5% |

Table 2 compares two different queuing approaches. The G/M/k results are for an empirical interarrival times, exponential service times (mean 2 days) and k parallel servers. For validation purposes we would like to compare the G/M/k results to M/M/k[15] results (similar to [4]) and so, even though the interarrival distribution has a fatter left tail than the exponential distribution, we nevertheless make an exponential fit. The resulting waiting times in the queue are presented in Table 2 and show that the (invalid) M/M/k assumption grossly underestimates these times. This observation is consistent with the fact that the exponential model underestimates the probability of short interarrival times. Note that the total time to service completion (time in the queue plus time in the service) adds an average of 2 days to these results making them much more comparable for large k. The final column of Table 2 shows the percent of the time analysts are busy and comes from (23). This equation only includes mean results which, since the exponential fit does capture the mean of the data perfectly (unlike the higher moments of the distribution), is the same across the two queuing models.

## 5.3  Threats to Validity

The under-reporting of problems by customers can skew the dataset making the right tail of the $D_i$ distribution heavier than it appears. Three main types of defects are not reported to the service desk: 1) defects with low severity with obvious workarounds; 2) non-reproducible defects that the customers encounter during coincidences of multiple events which disappear after restarting the software; and 3) rediscoveries of know defects that were not fixed the last time.

Underreporting may bias the analysis of actual software quality (Section 5.2.1). However, bias will be consistent across releases as long as underreporting is. Underreporting will not affect resource allocation processes (Section 5.2.2), since service and maintenance teams are interested in prediction of the actual number of support or special build requests. For them a bug that is not reported does not exist.

Generalizing our findings from a single-case design study to all situations is obviously not possible [18]. However, this design is based on the rationale of the *critical case* [18], confirming the applicability of our metrics to a large enterprise software. Also, it is not straightforward to perform a comparative study with existing metrics. Our metrics $M_{1-6}$ are complementary to but different from existing metrics making direct comparisons impossible. $M_7$-related queuing models have been used in the past [4]. However, the shape of the inter-arrival times distribution prohibits application of the queuing model used in [4] to our dataset: the M/M/k values underestimates waiting time (see discussion in Section 5.2.2).

## 6  Conclusions

Defect rediscovery is an important problem affecting both software manufacturers and customers. We have introduced a set of practical metrics designed to assess risks associated with defect rediscovery. The metrics can help the QA team with performance analysis of QA processes. They aid support and maintenance teams with resource allocation and with estimation of risk associated with under-staffing. Finally, the metrics provide customers with information on quality of various software products to help identify products best suited for their needs. The metrics can be applied to any defect rediscovery dataset and other distributions than those used here. We believe that these metrics are applicable to other software products. We also presented a validation case study showing application of the metrics to industrial data.

---

[15] Exponential interarrival times ($\lambda = 982$), exponential service times (mean 2 days) and k parallel servers.

# Appendix A Distribution Analysis

In order to find an analytic distribution that would be able to fit each of the yearly datasets, we use an L-moments ratio diagram [16]. This diagram is a goodness-of-fit tool to determine the probability distribution of the data. The L-moments are chosen since they are less biased and are less sensitive to outliers than ordinary moments [16], [19].

The diagram is shown in Figure 8 and the hollow circles denote each of the yearly datasets of $D_i$. The diagram shows the fits of the following widely used distributions [20]: Exponential (EXP), Normal (NOR), Gamma (GUM), Rayleigh (RAY), Uniform (UNI), Generalized Extreme Value (GEV), Generalized Logistic (GLO), Generalized Normal (GNO), Generalized Pareto (GPA), generalization of the Power Law, Pearson Type III (PE3), and Kappa (KAP). The diagram shows that the data is best approximated by a Kappa distribution as all data lie in the Kappa applicability space[16], with the Pearson Type III distribution the second best choice (data points lie around PE3 L-moments ratio line).

The diagram is shown in Figure 8. The hollow circles denote each of the 14 yearly datasets of $D_i$. The diagram shows the fits of the following widely used distributions [20]: Exponential (EXP), Normal (NOR), Gamma (GUM), Rayleigh (RAY), Uniform (UNI), Generalized Extreme Value (GEV), Generalized Logistic (GLO), Generalized Normal (GNO), Generalized Pareto (GPA), generalization of the Power Law, Pearson Type III (PE3), and Kappa (KAP). The lines and points on the legend depict possible values of L-skewness and L-kurtosis of these distributions with the exception of the Kappa distribution (represented by a plane). The closer the hollow circles to a given distribution point or line the better the fit of a given distribution.

Note that distributions with a small number of parameters (EXP, NOR, GUM, RAY, and UNI) show up as points and are clearly distant from any of the data points, therefore are not good fits. The applicability spaces for distributions with a larger number of parameters are represented by lines. Of these curves only the PE3 distribution appears to fit the L-moments well. The applicability space of KAP distribution (with one additional degree of freedom to the PE3) is represented by a plane bounded by GLO line above and "Theoretical limits" line below; hence each of the data points can be modeled by KAP distribution.

Table 3. AIC

| Distribution | v.1 | v.2 | v.3 | v.4 |
|---|---|---|---|---|
| PE3 | 12397 | 4481 | 2069 | 4797 |
| KAP | 11277 | 4309 | 2390 | 4304 |
| Compound KAP | 9392 | 4283 | 2934 | 4238 |

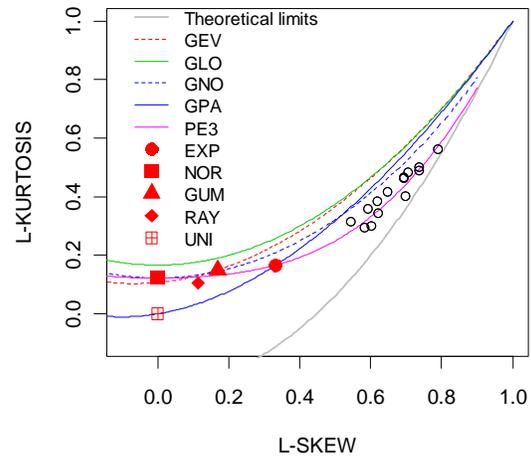

Figure 8. L-moments ratio diagram of $D_i$ for all releases per year (years 1 – 5). The hollow circles denote each of the yearly datasets of $D_i$. The diagram shows the fits of the following distributions: Exponential (EXP), Normal (NOR), Gamma (GUM), Rayleigh (RAY), Uniform (UNI), Generalized Extreme Value (GEV), Generalized Logistic (GLO), Generalized Normal (GNO), Generalized Pareto (GPA), generalization of the Power Law, Pearson Type III (PE3), and Kappa (KAP). The lines and points on the legend depict possible values of L-skewness and L-kurtosis of the distributions. Kappa (KAP) distribution L-skewness and L-kurtosis is a plane bounded by GLO distribution line above and the "Theoretical limits" line below and is not shown on the legend. Based on this figure, Kappa distribution is the only one that is applicable to modeling each of the datasets.

---

[16] The Kappa distribution's applicability space is a plane bounded by GLO and "Theoretical limit" L-moments ratio lines [16] on Figure 8.

The analysis procedure is adequately shown even if we limit the scope of the analysis to the four datasets of $D_i(1,2)$ showing rediscovery data for the second year of each release. We note that due to heavy tails, the exponential distribution does not provide an adequate fit to the data. Based on the data from the L-moments ratio diagram, we fit the data using the two best performers: Pearson Type III and Kappa distributions (remaining distributions provide inadequate fit). The QQ-plots showing goodness of fit are shown, accordingly, in Figure 9 and Figure 10. Based on Akaike's information criterion (AIC) [17] the Kappa distribution provides a better fit [17] than Pearson Type III for three datasets out of four (see Table 3).

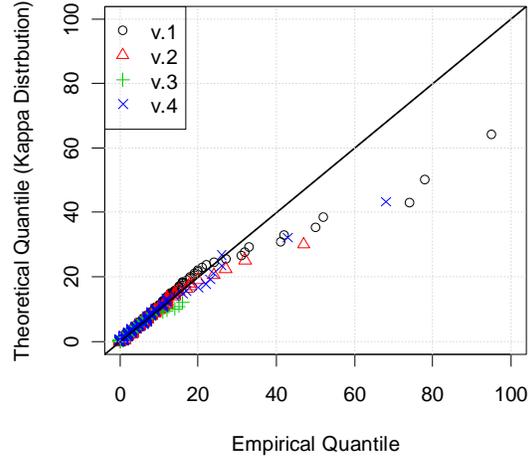

Figure 10. QQ plot of the empirical vs. KAP distributions' quantiles.

However, Figure 10 suggests that even the Kappa distribution isn't sufficiently flexible to fit both left and right tails of the empirical distribution. In order to overcome this obstacle we resort to a compound Kappa distribution[18].

The Kappa distribution [16] is a flexible 4-parameter distribution suited for fitting heavy-tail data. This distribution contains the Exponential, Weibull, Generalized Extreme Value, and Generalized Pareto distributions as special cases[19]. Its cdf is:

$$F(x) = \left\{ 1 - h\left[ 1 - \frac{\kappa(x-\xi)}{\alpha}\right]^{1/\kappa} \right\}^{1/h}. \quad (24)$$

The parameters $\xi$, $\alpha$, and $\kappa$ and $h$ describe location, scale, and shape, respectively. The associate quantile function is:

$$F^{-1}(u) = \xi + \frac{\alpha}{\kappa}\left[1 - \left(\frac{1-u^h}{h}\right)^{\kappa}\right]. \quad (25)$$

The first Kappa distribution with cdf $F_a$, fits the left tail of the dataset (in the range $[0, \rho]$) and the second with cdf $F_b$ fits the right tail in the range $(\rho, \infty)$. We select these partition points, $\rho$, for each of the four datasets by minimizing the sum of squares of the residuals between fitted and empirical data. For other techniques see [19]. Table 4 presents values of $\rho$. $F_a$ and $F_b$ are fitted independently; the resulting cumulative distribution function looks like:

$$F_c(d) = \begin{cases} ww_1 F_a(d), D \leq \rho \\ w[w_1 F_a(\rho) + w_2 F_b(d)], D > \rho \end{cases}, \quad (26)$$

where $w, w_1, w_2$ are the normalization constants

$$w_1 \equiv \text{ecdf}(\rho),$$
$$w_2 = 1 - w_1,$$
$$w = [w_1 F_a(\rho) + w_2 F_b(\infty)]^{-1} \quad (27)$$
$$= [w_1 F_a(\rho) + w_2]^{-1},$$

where ecdf is the empirical distribution function. We use the Weibull form [22] of the empirical distribution function: given a vector of observations $y$ sorted in ascending order, with sample size $n$, the unbiased non-exceedance probability of the $i$-th observation is given by:

$$\text{ecdf}(i) = i/(n+1). \quad (28)$$

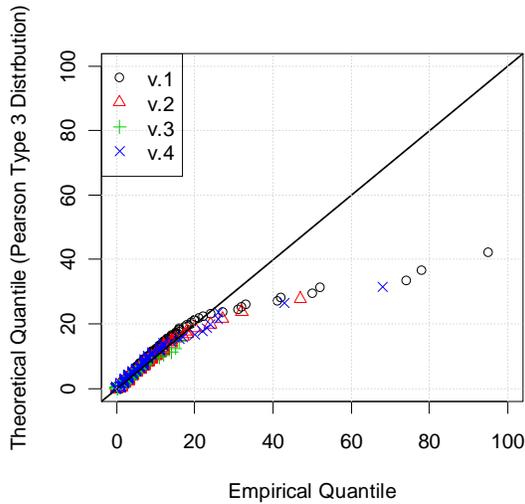

Figure 9. QQ plot of the empirical vs. PE3 distributions' quantiles.

---

[17] The lower the value of AIC – the better the fit.

[18] Compound distributions are used for modeling complex datasets in finance [21].
[19] Parameter values of distributions for our datasets suggest that we cannot "simplify" Kappa or Compound Kappa distributions.

The quantile function of the compound distribution can be obtained by inverting (26):

$$F_c^{-1}(u) = \xi_\omega + \frac{\alpha_\omega}{\kappa_\omega}\left[1 - \left(\frac{1-z(u)^{h_\omega}}{h_\omega}\right)^{\kappa_\omega}\right], \quad (29)$$

where

$$z(u) = \begin{cases} u/(ww_1), & u \leq ww_1 F_a(\rho) \\ [u - ww_1 F_a(\rho)]/ww_2, & u > ww_1 F_a(\rho) \end{cases}.$$

The parameters' lower index, $\omega$, specifies their affiliation with the first ($a$) or second ($b$) Kappa distributions; $\omega = a$ if $u \leq ww_1 F_a(\rho)$ and $\omega = b$ otherwise.

Table 4. Values of $\rho$

| Variable | v.1 | v.2 | v.3 | v.4 |
|---|---|---|---|---|
| $\rho$ | 15 | 15 | 8 | 10 |

The goodness of fit of the compound distribution is shown on QQ-plot in Figure 11. The QQ-plot suggests that the compound distribution provides a good fit to the underlying data. In addition, based on the AIC data given in Table 2, the compound Kappa distribution provides better fit than the Kappa distribution for three datasets out of four. This implies, from the AIC perspective, that increase of distribution complexity is offset by a significantly better fit. Unfortunately, we cannot use the Kolmogorov-Smirnov or chi-squared tests due to the large number of tied observations, as Figure 12 shows.

All of the distributions above are fitted to the data using the method of L-moments [16]. We have chosen this technique over the classical method of moments due to a more accurate estimate of the distribution's right tail [16], [19].

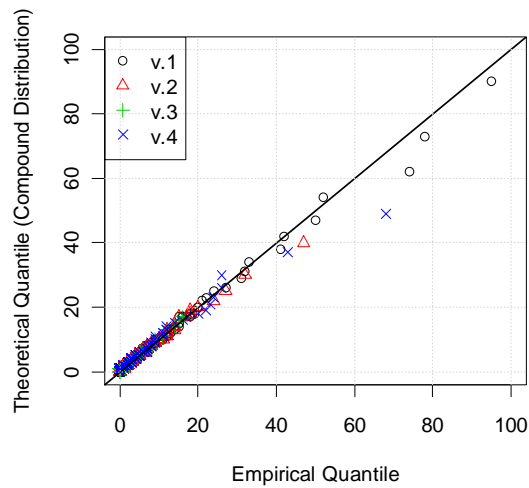

Figure 11. QQ plot of the empirical vs. Compound distributions' quantiles.

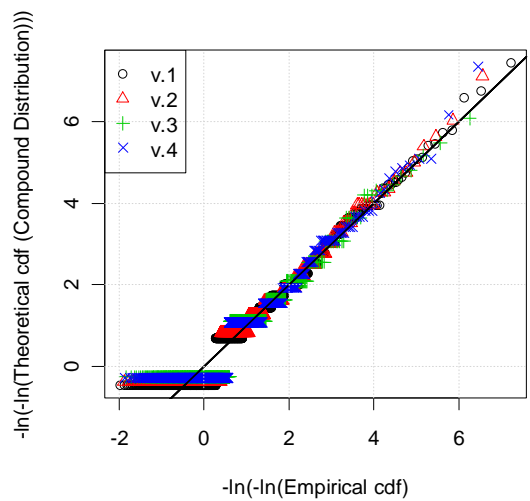

Figure 12. Plot of the empirical cdf vs. Compound Kappa distribution theoretic cdf.